\documentclass[12pt]{article}
\usepackage{amsfonts,amsbsy}
\newcommand{\be}{\begin{equation}}
\newcommand{\ee}{\end{equation}}

\newtheorem{definition}{Definition}

\newenvironment{statement}{\begin{itemize} \em}{\end{itemize}} 

\newcommand{\bea}{\begin{eqnarray}}
\newcommand{\eea}{\end{eqnarray}}
\newcommand{\In}{I}

\newcommand{\SPTLn}{\bar{{\cal S}}_L(n)}
\newcommand{\SPTA}{\bar{{\cal S}}}
\newcommand{\cTA}{{\cal T}(\mathbf{A})}
\newcommand{\mA}{\mathbf{A}}
\newcommand{\AT}{\tilde{\mathbf{A}}}
\newcommand{\VOL}{{\cal V}}
\input{epsf.tex}

\begin{document}

\vskip -4cm

\begin{flushright}
FTUAM-99-4

IFT-UAM/CSIC-99-5
\end{flushright}

\vskip 0.2cm

{\Large
\centerline{{\bf Combinatorics of lattice paths    }}
\centerline{{\bf   with and without spikes}}
\vskip 0.3cm

\centerline{\qquad  
  A. Gonz\'alez-Arroyo   }}
\vskip 0.3cm

\centerline{Departamento de F\'{\i}sica Te\'orica C-XI and}
\centerline{Instituto de  F\'{\i}sica Te\'orica C-XVI}
\centerline{Universidad Aut\'onoma de Madrid,}
\centerline{Cantoblanco, Madrid 28049, SPAIN.}
\vskip 10pt

\vskip 0.8cm

\begin{center}
{\bf ABSTRACT}
\end{center}

We derive a series of results on random walks on a d-dimensional hypercubic
lattice  (lattice paths). We introduce the notions of terse and simple paths
corresponding to the path having no backtracking parts (spikes). These paths 
label equivalence classes which allow a rearrangement of the sum over paths.
The basic combinatorial quantities of this construction are given. These 
formulas are useful when performing strong coupling (hopping parameter) 
expansions of lattice models. Some applications are described.  
\vskip 1.5 cm
\begin{flushleft}

\end{flushleft}

\newpage

\section{Introduction}
The strong coupling expansion is   a useful  analytical technique to study
lattice models.  In the context of lattice gauge theories it has been used
since early days to investigate the behaviour of the system far from the
continuum limit~\cite{Wilson1}. This technique is related to the high temperature
expansions of classical  Statistical Mechanics. In the case of matter
fields (continuous spin variables) with  nearest neighbor interactions the
technique involves a {\em hopping parameter expansion} giving rise to a
representation of the free energy and the propagators (correlation functions)
in terms of random walks (see Ref.~\cite{review} and references therein).
In its application to gauge theories the
contribution of each random walk includes  the corresponding Wilson loop.
Since the early calculations of this type~\cite{Kawam} the special behaviour
of {\em backtracking} paths was recognized. Backtracking occurs when
the walk makes two consecutive opposite  steps: the first in one direction and
the next one in  the reverse
direction.  Part of their special character
is related to  the fact  that the expectation value of a Wilson loop
is suppressed like an exponential of its area.  At infinitely strong coupling
and large $N$ this leaves loops which are {\em pure backtrackers}.  Actually, for unitary
gauge fields the backtracking part of a path is independent of the gauge
fields themselves. The problem of summing over these type of paths becomes
independent of the expectation value of gauge fields and is a pure
combinatorial problem. The problem was avoided, nevertheless, in the mentioned
lattice QCD strong coupling expansions~\cite{Kawam,Froh} by setting the Wilson
parameter $r$ equal to $1$. This kills backtracking paths from quark
propagators. Later, strong coupling calculations at $r \ne 1$ were performed by 
the effective potential method~\cite{EfPot} and its connection to the backtracking-path 
resummation problem was never established. Recently, motivated by the 
strong coupling expansion of supersymmetric Yang-Mills theory~\cite{GGAP}
we fell back into the problem. The effective potential method was
unavailable in this case and we attacked the problem of backtracking random
walks. The solution, presented here, gives for the lattice QCD case results in 
agreement with the effective potential method. We believe that the results  
and techniques can be of interest in other situations and hence we decided 
to collect proofs and results in this paper. 

This paper is written in a  self-contained form. In the next section we
introduce the basic notation and definitions. Not to conflict with other
definitions we will refer to random walks as {\em lattice paths}
and to backtracking parts as {\em spikes}. Then,  the main  result on
resummation over pure backtracking parts is presented.
Section 3 gives the expression of a kind of matrix generating functional
for paths with no spikes. This expression is useful for strong coupling expansions
of lattice gauge theories. In Section 4, we consider   closed paths. In this
case the notion of a {\em simple} path turns out to be useful. Formulas
similar to those given in the previous two sections are given for the case of
simple paths. Finally, in section 5 we give a few applications
which exemplify the way in which the previous results enter in physical expressions
within strong coupling expansions. This includes the pure spike contribution
to the free energy of a Gaussian model and the mesonic effective action.
This is the surviving part if the fields are coupled to a random $U(N)$
gauge field at large $N$~\cite{largeN}. The reader who is not interested in
proofs can jump directly to the last section. Our main results are given in
formulas~(\ref{expF}),~(\ref{resT}),~(\ref{resTp1}-\ref{resTp3}),~(\ref{FtildeRes}),
~(\ref{NtildeRes}),~(\ref{Ftildep1Res}) and (\ref{TTILDEres}).

\section{Reducing  paths}
In this section  we will introduce the basic notation and definitions.
We will be working in arbitrary space-time dimension d. Vector indices $\mu$ go from 
$0$ to $d-1$.  We will need to introduce an index set $\In$ with $2d$ elements. 
For every space-time direction $\mu$ there are two elements $\mu$ and
$\bar{{\mu}}$. They correspond to the two senses associated to each direction
(forward and backward).
Now consider our space-time lattice ${\cal L} \equiv \mathbf{Z}^d$. We might 
associate to any element $\alpha$ in $\In$ a lattice vector $V(\alpha)$ as follows:
$$ \mu \longrightarrow V(\mu)\equiv e^{(\mu)}\ \ \ \ \bar{\mu}
\longrightarrow  V(\bar{\mu})\equiv -e^{(\mu)}\
,$$
where $e^{(\mu)}$ is the unit vector in the $\mu$ direction.
Given one element $\alpha \in \In$ the element $\bar{\alpha}$ denotes the
reversely oriented one ($\bar{\bar{\mu}}=\mu$).

Now we proceed to give a few definitions:
\begin{definition}
A  {\bf lattice path} of length L is an element 
$\gamma \equiv (n,\vec{\alpha}) \in {\cal L} \times \In^L$. The point $n \in  {\cal L}$ is
the {\bf origin} of the path, and $\vec{\alpha}$ is the  {\bf path sequence}, specifying 
the steps to take to describe the path.   
\end{definition}
The {\bf endpoint} of a path $(n,\alpha_1,\ldots ,\alpha_L)$ is given  by the 
lattice point $m=n+V(\vec{\alpha})= n+\sum_{i=1}^{L}V(\alpha_i)$. We might now introduce the 
following nomenclature for the set of paths. Let ${\cal S}_L(n)$ be the space
of all paths with origin $n$ and length $L$.  ${\cal S}(n)$ labels the set 
of all paths with origin $n$ and any length. We might also 
fix the origin and endpoint and write  ${\cal S}_L(n\rightarrow m)$.

The total number of paths of length $L$ is easy to  count: $N(L)=(2d)^L$.
For length $L=0$ we will consider that there is a unique path with origin
in $n$, which we will call the {\bf path of zero length}. To any path
$\gamma\equiv(n,\vec{\alpha})$ of length   $L$, there corresponds a path
called its {\bf reverse path} of equal length and labeled
$\gamma^{-1}\equiv(m,\vec{\beta})$.
The origin of the reverse path $m$ is the endpoint of the original path and
vice versa. The path sequence is the reversely ordered one
($\beta_i=\bar{\alpha}_{L-i+1}$).  We will also introduce a path
composition operation. Given a path $\gamma\equiv(n,\vec{\alpha})$ whose
endpoint is $m$, and another path $\gamma'\equiv(m,\vec{\beta})$, we can
construct the composed path $\gamma \circ \gamma'\equiv
(n,\vec{\alpha},\vec{\beta})$.

Now we will give some more definitions. 
\begin{definition}
A path $\gamma \equiv (n,\alpha_1,\ldots ,\alpha_L)$ has {\bf spikes}
if there exist one integer $i$  ($1 \le i \le (L-1)$) such that
$\alpha_{i+1}=\bar{\alpha_i}$. In the converse case one says that the path 
is {\bf terse} or has   {\bf no spikes}. The set of all paths without
spikes(terse)
of length $L$ and  origin $n$  is labeled  $\bar{{\cal S}}_L(n)$ 
(  $\bar{{\cal S}}_L(n\rightarrow m)$ if the endpoint is fixed to $m$).
\end{definition}
It is not difficult to obtain $\bar{N}(L)$: the number of elements
of $\SPTLn$. Its value is $2 d (2d-1)^{L-1}$ for $L \ge 1$. The path of zero
length is terse $\bar{N}(0)=1$.

Now we will classify the set of paths into subsets labeled by a 
terse path. Let us first present  the results:

\begin{statement}

\item There exist a projection 
$\pi: {\cal S}(n\rightarrow m) \longrightarrow
\SPTA(n\rightarrow m) \subset  {\cal S}(n\rightarrow m)$
such that to every path $\gamma$ it associates  a terse path  
$\pi(\gamma)$, called  its {\bf reduced path}. If the  length of 
$\gamma$ is $L$, then the 
length of $\pi(\gamma)$ is $L-2p$, for some integer $p$.
 
\end{statement}

\begin {definition} 
If $\pi(\gamma)$ is the path of length zero, then $\gamma$ 
is said to be a  {\bf pure spike} path.
\end{definition}

The construction of  $\pi(\gamma)$ proceeds iteratively.
If the path  $\gamma=(n,\vec{\alpha})$ is terse, 
then $\pi(\gamma) = \gamma$. 
Otherwise, one can start to scan the sequence
of indices $i$ in increasing order, until one finds a value of 
$i$ such that $\alpha_{i+1}=\bar{\alpha_i}$. This by hypothesis must hold
for some $i$. Then, one can eliminate the elements $i$ and $i+1$ from the 
sequence, thus defining a new path of length $L-2$. Then, one can apply 
the procedure once more to the resulting path. In this way, one must 
proceed iteratively until the iteration terminates.  This   must necessarily
happen since the length of the original path $L$ is  finite. 
The iteration can terminate in two ways. Most frequently, one would reach, 
at some stage of the iterative procedure, a path without spikes. Then this 
is precisely $\pi(\gamma)$. In some cases, the iteration proceeds until 
there are no more elements left in the sequence $\vec{\alpha}$. In this 
case we  would say that the corresponding reduced path   is the 
path of zero length, and $\gamma$ is a pure spike.

The sets $\pi^{-1}(\hat{\gamma})$ will play an important role in 
our  construction. Our main interest  is to determine the 
numbers $N(\hat{\gamma},p)$: the number of paths of length 
$2p+\bar{L}$ whose reduced path is $\hat{\gamma}$ (whose length is 
$\bar{L}$). For the construction we will need to introduce two groups of
operations on the sets of paths:
\bea 
&\phi: {\cal S}_L(n) \longrightarrow   {\cal S}_{L-1}(n)\\
\nonumber
& \mbox{such that  for }   \gamma=(n,\alpha_1, \ldots \alpha_L) \ \ \mbox{, we have } 
\phi(\gamma)=(n,\alpha_1, \ldots \alpha_{L-1})\\
&
\phi_{\alpha}: {\cal S}_L(n) \longrightarrow   {\cal S}_{L+1}(n)\\
&\nonumber \mbox{ For } \ \gamma=(n,\alpha_1, \ldots \alpha_L)\ \mbox{ and } \alpha
\in I  \mbox{, we have }    \phi_{\alpha}(\gamma)=(n,\alpha_1, \ldots
\alpha_{L},\alpha)
\eea

What we need to know is what is the interplay between these operations  and
the projection $\pi$. Let us consider a path $\gamma=(n,\alpha_1, \ldots
\alpha_L)$ whose reduced path is $\pi(\gamma)=(n, \beta_1, \ldots ,
\beta_{\bar{L}})$. We are interested in the reduced path
$\pi(\phi_{\alpha}(\gamma))$. By the iterative definition of $\pi$, we see that 
after some iterations we would end up with a path
$\phi_{\alpha}(\pi(\gamma))$. Now there can be two cases: if $\bar{\alpha} \ne
\beta_{\bar{L}}$ this path is terse and hence
$\pi(\phi_{\alpha}(\gamma))=(n, \beta_1, \ldots
,\beta_{\bar{L}},\alpha)=\phi_{\alpha}(\pi(\gamma))$;
for the special case  $\bar{\alpha} = \beta_{\bar{L}}$, one must still apply 
one reduction step and the result is $\pi(\phi_{\beta_{\bar{L}}}(\gamma))=
(n, \beta_1, \ldots ,\beta_{\bar{L}-1})=\phi(\pi(\gamma))
$. Now we study
$\pi(\phi(\gamma))$. There are again two cases: if $\alpha_L=\beta_{\bar{L}}$ 
the result is $(n, \beta_1, \ldots ,\beta_{\bar{L}-1})=\phi(\pi(\gamma))$; in the rest of cases
we have  $(n, \beta_1, \ldots
,\beta_{\bar{L}},\bar{\alpha}_L)=\phi_{\bar{\alpha}_L}(\pi(\gamma))$. These
results can be proven in a similar way as  for $\phi_{\alpha}$. 

Now we will make use of the previous results. Consider a terse path
$\hat{\gamma}=(n,  \beta_1, \ldots , \beta_{\bar{L}})$ of non-zero length
$\bar{L}$, and consider the set ${\cal S}(\hat{\gamma},p)$ of all paths
$\gamma$ of length $L=\bar{L}+2p$, with $p \ge 1$ an integer,  whose reduced
path is $\hat{\gamma}$. Then we can
conclude:
\begin{statement}
\item The application $\phi$ induces a mapping from  \\ ${\cal
S}(\hat{\gamma},p)$ into  ${\cal S}(\phi(\hat{\gamma}),p)\cup_{\alpha \ne
\beta_{\bar{L}}} {\cal S}(\phi_{\bar{\alpha}}(\hat{\gamma}),p-1)$, which is 
bijective.
\item Henceforth, the number of paths  $N(\hat{\gamma},p)$ in
${\cal S}(\hat{\gamma},p)$ satisfies:
\[N(\hat{\gamma},p)=N(\phi(\hat{\gamma}),p)+ \sum_{\alpha \ne \beta_{\bar{L}}} 
N(\phi_{\bar{\alpha}}(\hat{\gamma}),p-1)\quad . \]
 Actually, the number  $N(\hat{\gamma},p)$ does 
not depend on the path $\hat{\gamma}$ but only on its length $\bar{L}$. We
thus conclude:
\be
\label{rel1}
N(\bar{L},p)=N(\bar{L}-1,p)+(2d-1)\, N(\bar{L}+1,p-1)\ \ .
\ee
\item For a pure spike path $(n,\alpha_1, \ldots ,\alpha_L)$,  we might apply  $\phi$ and produce a path of
length $L-1$ with reduced path $(n,\bar{\alpha}_L)$. This is also bijective
and leads to:
\be
\label{rel2}
N(0,p)=2d\,  N(1,p-1) .
\ee
\end{statement}

The proof of the previous statements is as follows. The bijectivity can be
shown by the existence of an inverse transformation. This is basically
$\phi_{\alpha}$ with $\alpha$ chosen appropriately. To prove  that
$N(\hat{\gamma},p)$ only depends on the length can be done by induction.
Prove directly by construction that the statement is true for paths of short length
(it is easy to solve the problem up to   $L=4$ for example). Then one
assumes that the statement is verified  up to  length $L=\bar{L}+2p$ (for
any $p$). Then one can use the formulas (\ref{rel1}) and (\ref{rel2}) to show
that the statement is true  for paths of length $L+1$. Notice that if the right
hand side does not depend on the actual terse  paths but only on its
lengths, and since these  lengths only depend on the length of the reduced path
$\hat{\gamma}$ of the left hand side, the result follows.

By repeated application of the  relations Eq. (\ref{rel1}) and (\ref{rel2}),
together with the initial condition  $N(\bar{L},0)=1$, one can obtain all the 
$N(\bar{L},p)$ values. To exploit  these  relations we will 
introduce  the following generating functions:
\bea
F(\bar{L},z)= \sum_{p=0}^{\infty}\, z^p\, N(\bar{L},p)\\
G(y,z)= \sum_{\bar{L}=0}^{\infty}\, F(\bar{L},z)\, y^{\bar{L}}
\eea
  Multiplying the relation~(\ref{rel1}) by the appropriate 
  powers of $z$ and $y$ and summing over $p$ and $\bar{L}$, one gets:
  \be
  G(y,z)-F(0,z)= y\, G(y,z) +\frac{(2d-1)z}{y}(G(y,z)-F(0,z)-yF(1,z))
  \ee
and from it, one can  write $G$ in terms of $F$:
 \be
 \label{expG}
 G(y,z)=\frac{1}{(y (1-y) - (2d-1)z)} \left( \left(\frac{y}{2d} -
(2d-1)z\right)F(0,z) +
 y \frac{2d-1}{d}\right)
  \ee
Notice that the zeroes of the denominator in the previous expression can give rise to
singularities,  even for small values of $z$ and $y$, unless the numerator
vanishes at these zeroes. This must actually happen since  $G$ and
$F$ can be shown to be analytic in a neighborhood of $y=z=0$ ( This follows
from $N(\bar{L},p) < (2d)^{\bar{L}+2p}$). This allows one to determine
$F(0,z)$: 
\be 
F(0,z)= \frac{2d-1}{d} \frac{1}{1 +\frac{d}{d-1}\sqrt{1-4 (2d-1) z}}. 
\ee
 Now plugging this expression into~(\ref{expG}) we obtain the formula
 for $G(y,z)$. 
 
 The expressions  can be simplified with a suitable change 
 of variables. Let us introduce the variable $\xi$:
 \bea
 \label{xidef}
 \xi(z)=\frac{1-\sqrt{1-4(2d-1)z}}{2}\\
\mbox{ with inverse }  z(\xi)=\frac{\xi(1-\xi)}{2d-1}
 \eea
 Then one can conclude:
 \bea
 F(0,z(\xi))= \frac{1}{1-\frac{2d}{2d-1}\xi}\\
 G(y,z(\xi))= - \frac{(2d-1)(1-\xi)}{2d-1-2d\xi}\  \frac{1}{y+\xi-1} 
 \eea
 Notice that the only dependence on $y$ sits in the last denominator. 
 It is now fairly simple to obtain $F(\bar{L},z)$ by picking the relevant
 power of $y$ in the expansion. One gets:
 \be
 \label{expF}
 F(\bar{L},z(\xi))=\frac{1}{(1-\frac{2d}{2d-1}\xi)}
\frac{1}{(1-\xi)^{\bar{L}}}
 \ee
The last formula is the main one of this chapter.  From it one can obtain the numbers
$N(\bar{L},p)$, by differentiation or Cauchy integration. This we will do
later.

Before, as a check of our formulas, one can compute the number of  paths of length 
$L$ as a sum over the number of  terse paths  times the number of 
paths of length $L$ having a given terse path as  reduced path:
\be
(2d)^L=N(L)= \sum_{p=0}^{[\frac{L}{2}]}\, \bar{N}(L-2p)\, N(L-2p,p)
\ee
To check all formulas at the same time we can multiply the expression by 
$z^L$ and sum over $L$.  We get:
\bea
\nonumber
&\frac{1}{1-2dz}= F(0,z^2) +\sum_{\bar{L}=1}^{\infty}  z^{\bar{L}}\,
F(\bar{L},z^2)\,   2 d\,
(2d-1)^{\bar{L}-1}=\\ & \frac{1}{(1-\frac{2d}{2d-1}\xi(z^2))}\, (
1+\frac{2dz}{1-\xi(z^2)-(2d-1)z
})=\\
\nonumber 
&\frac{1-\xi(z^2)+z}{(1-\frac{2d}{2d-1}\xi(z^2))(1-\xi(z^2)-(2d-1)z)} \quad .
\eea
For the third identity of the  previous formula we have  resummed a
geometric series. Finally, to prove that the right hand side of the previous
equation coincides with the left hand side, one must simply manipulate
algebraically the expression and use the relation $\xi(z^2) (1-\xi(z^2))
=(2d-1) z^2$.

We conclude this section by extracting the numbers $N(L,p)$ themselves. This
can be done by employing the  expression of the generating function
(\ref{expF}), and making a contour integral in the complex plane of $z$
around  the origin, and using Cauchy's theorem. It is more practical to
change variables from $z$ to $\xi$ in the integral. Notice that for
$|z|$ small enough, the contour in $\xi$ also encircles the origin and the
function $F(L,z(\xi))$ has no singularities inside. The resulting integrand
is a product of negative powers of $\xi$ and of $(1-\xi)$,  times the factor
$1/(1-\frac{2d}{2d-1}\xi)$ coming from (\ref{expF}). If one expands this
denominator in powers of $\xi$, it is not hard to show that:
\be
\label{expNLp}
N(L,p)= \sum_{j=0}^{p} (2d)^j\, (2d-1)^{p-j}\, \frac{(L+2p-j-1)!}{(p-j)!\,
(L+p)!}\, (L+j)\ . 
\ee
We see that the resulting expression is a polynomial in $d$ of degree $p$.
To obtain the coefficients of the different powers of $d$, one could expand
the power of $(2d-1)$ in powers of $d$ and  rearrange the summation. The
method can be carried out  but is fairly lengthy and complicated. A short cut to arrive to
the same final expression is to multiply and divide Eq.~(\ref{expNLp}) by $j!$.
Then one replaces the factors  $j!$ and $(L+2p-j-1)!$ by their standard
integral representation(that of Euler's gamma function) and performs the sum
in $j$. The expression is then given as a double integral over two
variables $\alpha$ and $\beta$ going from $0$ to $\infty$. Now one can
perform the standard trick in computing Feynman integrals by changing
variables to $\lambda \equiv (\alpha + \beta)$ and the {\em Feynman parameter}
$x \equiv \frac{\alpha}{\lambda}$. The integration over $\lambda$ can be performed and we
arrive at: 
\bea
\nonumber
&N(L,p)=\left(\begin{array}{cc}L+2p\\ p \end{array} \right)\, \int_0^1 dx\,
x^{L+p-1}\, (2d-x)^{p-1}\,  ( L(2d-x) + 2dp(1-x))=\\ 
\label{Nexp2}
&=\, \left(\begin{array}{cc}L+2p\\ p \end{array} \right)\, \sum_{s=0}^p
\frac{(2d)^s (-1)^{p-s}}{(L+2p-s)} \left(\begin{array}{cc}p\\ s
\end{array} \right)\, \left( L+\frac{s}{L+2p-s+1}\right)\quad .
\eea
The first equality in the previous expression is a Feynman parameter
integral representation of the numbers $N(L,p)$. The second one is
a representation as a polynomial in $d$, and it can be obtained easily from
the other. We have given the expression for $N(L,p)$
for completeness, though in actual applications it is more useful to work
with  the  generating function $F(L,z)$.

\section{Summing over terse paths}
In this section we will compute a matrix generating function for the set of
terse paths $\SPTLn$. This generating function turns out to be useful in applications to
strong coupling expansions of lattice models.
 Let us introduce a collection of matrices $\mathbf{A}_{\alpha}$
for $\alpha \in \In$. The interesting quantity to study is: 
\be
\label{defT}
{\cal T}(\mathbf{A}) = \sum_{L=0}^{\infty} \ \ \sum_{(n,\vec{\alpha})\in
\bar{\cal S}_L(n)}
\mathbf{A}_{\alpha_1} \cdots  \mathbf{A}_{\alpha_{L}} .
\ee
We will first   compute ${\cal T}(\mathbf{A})$ for matrices 
$\mathbf{A}$ satisfying  $\mathbf{A}_{\alpha} \mathbf{A}_{\bar{\alpha}}=
\lambda\, \mathbf{I}$ ({\em i.e.} a multiple of the identity). This condition 
is  satisfied in some of the most important applications of the formula. 
At the end of the section we will give the more general formulas.

To  facilitate the reading for those who are mostly interested in the result,
we will begin by giving the answer:
\bea
\label{resT}
&{\cal T}(\mathbf{A}) =(1-\lambda)(1+(2d-1)\lambda - \tilde{\mathbf{A}})^{-1}\\
&\mbox{with } \quad \tilde{\mathbf{A}}=\sum_{\alpha\in I} \mathbf{A}_{\alpha} 
\eea
The conditions on the  matrices  $\mA$ for which the previous expression
applies can be read out from it. The eigenvalues of $\AT$ must be small
enough for the inverse matrix entering in Eq.~(\ref{resT})  to exist. 
In the following paragraphs  we will give the proof of this result.

We begin by considering the set of all terse paths, with origin in $n$,
 length $L\ge 1$ and ending with step $\alpha$: $\bar{{\cal
S}}^{\alpha}_{L}(n)$. Applying a similar definition as Eq.~(\ref{defT})
to this set we get:
\be\label{defTP}
{\cal T}_{\alpha}(L,\mathbf{A}) =  \sum_{\gamma \in
\bar{\cal S}_L^{\alpha}(n)}
\mathbf{A}_{\alpha_1} \cdots  \mathbf{A}_{\alpha} .
\ee
Now, clearly the path $\phi(\gamma)$ is also a terse path and has    length 
$L-1$, but it cannot end with step $\bar{\alpha}$.  Hence:
\be
\label{equaT1}
{\cal T}_{\alpha}(L+1,\mathbf{A}) = \sum_{\beta \ne \bar{\alpha}} 
{\cal T}_{\beta}(L,\mathbf{A})\ \mathbf{A}_{\alpha}.
\ee
The formula is valid for $L \ge 1$. 

Now from it we will  derive an equation for 
${\cal T}_{\alpha}(\mathbf{A}) \equiv \sum_{L=1}^{\infty} {\cal
T}_{\alpha}(L,\mathbf{A})$.  Then our main quantity  ${\cal T}(\mathbf{A})$ 
is given by $\mathbf{I}+\sum_{\alpha\in I} {\cal
T}_{\alpha}(\mathbf{A})$. Summing Eq.~(\ref{equaT1}) over $L$ one gets:
\be
{\cal T}_{\alpha}(\mathbf{A})=\left({\cal T}(\mathbf{A})-
{\cal T}_{\bar{\alpha}}(\mathbf{A})\right)\mathbf{A}_{\alpha}
\ee
 Given ${\cal T}(\mathbf{A})$, these  are coupled equations  for the indices 
 $\alpha$ and $\bar{\alpha}$. We then write them as a single vector equation:
 \be
 \left({\cal T}_{\alpha}(\mathbf{A}),{\cal T}_{\bar{\alpha}}(\mathbf{A})\right)
\mathbf{\cal H} = {\cal T}(\mathbf{A})\, \left(\mathbf{A}_{\alpha},
\mathbf{A}_{\bar{\alpha}}\right)    
 \ee
 where  $\mathbf{\cal H}$ is an invertible matrix. This matrix and its
inverse are given by  the formulas:
\bea
&\mathbf{\cal H}=  \left( \begin{array}{cc} \mathbf{I} &
\mathbf{A}_{\bar{\alpha}}\\ \mathbf{A}_{\alpha} & \mathbf{I}
\end{array}\right)\\ 
&\mathbf{\cal H}^{-1}= \frac{1}{(1-\lambda)} \left( \begin{array}{cc}
\mathbf{I} & -\mathbf{A}_{\bar{\alpha}}\\ -\mathbf{A}_{\alpha} & \mathbf{I} 
\end{array}\right)
\eea
Then, for fixed  ${\cal T}(\mathbf{A})$, one can solve for ${\cal
T}_{\alpha}(\mathbf{A})$, obtaining:
\be
{\cal T}_{\alpha}(\mathbf{A})=\frac{{\cal T}(\mathbf{A})}{(1-\lambda)}
\, (\mathbf{A}_{\alpha} -\lambda)\ .
\ee
Finally summing both sides of the equation over $\alpha$  we get:
\be
{\cal T}(\mathbf{A})-\mathbf{I} =\frac{{\cal T}(\mathbf{A})}{(1-\lambda)}
\, (\tilde{\mathbf{A}}-2d \lambda)\quad  .
\ee
From this equation one can solve for  ${\cal T}(\mathbf{A})$ obtaining the 
formula (\ref{resT}).

We can now, as in the previous section, check the  formula by using it in
deriving a known result. Consider the sum over all paths
$\gamma=(n,\vec{\alpha})$ of the ordered product of the matrices
$\mathbf{A}_{\alpha}$. Since this is a geometric series it is easily resummed
to $(\mathbf{I}-\tilde{\mathbf{A}})^{-1}$. Now this result has to be
reobtained by splitting the sum over paths into a sum over terse  paths 
 and a sum over paths whose reduced path  is a given terse path. 
 In this way one is making use of the results of the last section and
of this section at the same time. The last summation can be performed in
terms of the generating function studied in the last section. One has:
\be
 \sum_{L=0}^{\infty} \ 
\sum_{(n,\vec{\alpha})\in \bar{\cal S}_L(n)}
\mathbf{A}_{\alpha_1} \cdots  \mathbf{A}_{\alpha_{L}}\,  F(L,\lambda)\ \ .
\ee
Now, given the form of $F(L,\lambda)$ given in expr.~(\ref{expF}), one
recognizes the structure given in Eq.~(\ref{defT}) with the
$\mathbf{A}_{\alpha}$ rescaled. We get:
\be  
\frac{1}{(1-\frac{2d}{2d-1}\xi(\lambda))}\, {\cal T}(\mathbf{A}/(1-\xi(\lambda)))
\ee
Now using our expression for $\cTA$ (Eq.~(\ref{resT})) and the relation
between $\xi(\lambda)$ and $\lambda$ (Eq.~\ref{xidef}) one  obtains the known
result.

Now, as announced at the beginning of the section we will give the result
without imposing  the condition $\mathbf{A}_{\alpha}
\mathbf{A}_{\bar{\alpha}}=\lambda\, \mathbf{I}$. It is not difficult by
following the same steps as before to show that in the general case we 
have:
\bea
\label{resTp1}
& {\cal T}(\mathbf{A})=(1-\tilde{\mathbf{B}})^{-1}\\
\label{resTp2}
& {\cal T}_{\alpha}(\mathbf{A})= {\cal T}(\mathbf{A})\, ( \mathbf{A}_{\alpha}-
\mathbf{A}_{\bar{\alpha}} \mathbf{A}_{\alpha})\, (\mathbf{I} -
\mathbf{A}_{\bar{\alpha}} \mathbf{A}_{\alpha})^{-1}\\
\label{resTp3}
& \mbox{    where    }  \tilde{\mathbf{B}} = \sum_{\alpha}\,  (
\mathbf{A}_{\alpha}-
\mathbf{A}_{\bar{\alpha}} \mathbf{A}_{\alpha})\, (\mathbf{I} -
\mathbf{A}_{\bar{\alpha}} \mathbf{A}_{\alpha})^{-1}
\eea
The previously given  formula~(\ref{resT}) follows as a special case from this one. 

To conclude this  section, we comment that, using the above formulas,
one can derive matrix generating
functions  for the sum of terse paths with fixed origin and endpoint.
For that purpose one simply has to
multiply $\mathbf{A}_{\alpha}$ by a phase $e^{\imath \varphi_{\alpha}}$. For
the reverse direction $\bar{\alpha}$ one has the complex conjugate phase
($\varphi_{\bar{\alpha}}=-\varphi_{\alpha}$),  so that the
condition $\mathbf{A}_{\alpha} \mathbf{A}_{\bar{\alpha}}=\lambda\,  \mathbf{I}$ is respected.  With a suitable
integration over the phases $\varphi_{\mu}$ one can restrict the sum over
terse paths  to those having fixed  endpoint. For example, if one wants to evaluate the
contribution to ${\cal T}(\mathbf{A})$ from paths whose origin is $n$ and
endpoint is $m$ one can simply write:
\be
\prod_{\mu} \left( \int_0^{2 \pi} \frac{d \varphi_{\mu}}{2 \pi}\,
e^{\imath\varphi_{\mu}(n_{\mu}-m_{\mu})}\right)\, {\cal
T}(e^{\imath \varphi_{\alpha}}\,\mathbf{A})\quad .
\ee

\section{Closed and Simple paths}
In this section we will look at closed paths: a path such that
its origin and endpoint coincide. One  frequently encounters
situations in which closed paths play an important role, such as in
computing the free energy or the fermion determinant. In those cases, for
every such path one has to evaluate a trace. For that purpose,  the notion of
terse paths is somehow insufficient. We would like to single out 
those paths $(n,\vec{\alpha})$ for which the last step $\alpha_L$ differs
from  $\bar{\alpha}_1$. We will call those paths simple. Let us now give more
precisely  the definition.
\begin{definition}
A  path $\gamma=(n,\vec{\alpha})$ of length $L$ is  {\bf simple}, if it is terse (without spikes) 
 and in addition one has  $\alpha_{L}  \ne \bar{\alpha}_1$.
\end{definition}
Simple paths can be open or closed, however their usefulness appears normally
when they are closed. The set of simple paths of length $L$ and  origin in $n$
is labeled  $\tilde{\cal S}_L(n)$. The set  $\tilde{\cal S}_0(n)$ is  given
by the path of zero length. 

By a similar procedure to  the one followed in section 1, one can associate
to any path $\gamma$ a given simple path $\tilde{\pi}(\gamma)$. To
construct  $\tilde{\pi}(\gamma)$ one starts by obtaining the reduced path
($\pi(\gamma)=\hat{\gamma}\equiv (n,\vec{\beta})$) 
associated to $\gamma$. Let us consider that its length is $L$ and
its origin  $n$. If this terse  path is simple, then  this is
precisely  $\tilde{\pi}(\gamma)$.  If not,  it is  due to 
$\beta_L$ being equal to  $\bar{\beta}_1$. Hence, we  eliminate the
first and last steps in the  sequence ( $\beta_1$ and $\beta_L$). The
resulting   path is terse and has length $L-2$. Notice, however,
that its origin is now $n+V(\beta_1)$, and not $n$. If the path is simple,
then it coincides with $\tilde{\pi}(\gamma)$, otherwise one has to repeat
the procedure once more. Eventually, one reaches a simple path, which could 
be just a path of zero length.

Our first goal is to develop similar counting rules for simple  paths as
those  obtained in section 2 for terse paths. In particular we
are  interested in computing the numbers
$\widetilde{N}(\tilde{\gamma},p)$: the number of paths $\gamma$ of length $l +2 p$ whose
associated simple path  is $\tilde{\pi}(\gamma)=\tilde{\gamma}$ (of length $l$). 
The generating function of these
numbers is:
\be
\widetilde{F}(\tilde{\gamma},z)= \sum_{p=0}^{\infty} z^p\, 
\widetilde{N}(\tilde{\gamma},p)\quad .
\ee
In what follows we will compute  this generating function and  the numbers
$\widetilde{N}(\tilde{\gamma},p)$.

The procedure that we will employ  is to relate these numbers with
$N(l,p)$.  For that purpose consider a path
$\gamma$
with $\tilde{\pi}(\gamma)=\tilde{\gamma}\equiv(m,\vec{\alpha})$  and consider its reduced path
$\pi(\gamma)\equiv(n,\vec{\beta})$. It is clear from the description of the
construction of $\tilde{\pi}(\gamma)$ that the path $\pi(\gamma)$ must be
the composition of three paths: 
\[ \pi(\gamma) = s \circ \tilde{\gamma} \circ s^{-1}                  \quad .  \]
The path $s\equiv(n,\vec{\rho})$ is a terse path of  length $p'$
($0\le p' \le p$) going from $n$ to $m$. If the path $\tilde{\gamma}$ has
length $l$ then  $\rho_{p'} \ne \bar{\alpha}_1 ,
\alpha_l $. This must hold, since  the composition $ s \circ
\tilde{\gamma} \circ s^{-1}$ must be terse. Conversely all paths $\gamma$ having
$\pi(\gamma) = s \circ \tilde{\gamma} \circ s^{-1}$ have $\tilde{\gamma}$ as
its associated simple path. Hence,  all one needs to  do is to count for each
case of $\pi(\gamma)$ the number of paths $\gamma$ of length $l+2p$. The main
formula is:
\be
\label{main2}
\widetilde{N}(\tilde{\gamma},p)= N(l,p) + \sum_{p'=1}^{p} N(l+2p',p-p')\,
(2d-2)
(2d-1)^{p'-1}  
\ee
The quantity  $(2d-2) (2d-1)^{p'-1}$ counts the number of acceptable terse paths
$s$  of length $p'$ going from any point $n$ to $m$. The word
acceptable refers to the condition $\rho_{p'} \ne \bar{\alpha}_1 ,
\alpha_l $. The previous formula~(\ref{main2}) is valid for $l$ and $p$
strictly positive. If any of the two is zero then $\widetilde{N}=N$.

A first conclusion from formula~(\ref{main2}) is that
$\widetilde{N}(\tilde{\gamma},p)$ only depends on the length $l$ of the
simple path $\tilde{\gamma}$. If we multiply both sides of equation by $z^p$
and sum over $p$, we get (for $l > 0$):
\be
\widetilde{F}(l,z)= F(l,z) + \sum_{p'=1}^{\infty}(2d-2) (2d-1)^{p'-1} z^{p'}
\ee
Now using the form of $F(l,z)$ one gets:
\be
\label{FtildeRes}
\widetilde{F}(l,z)= \frac{1}{1-2 \xi(z)}\, \frac{1}{(1-\xi(z))^l}\quad . 
\ee
This formula is valid for $l > 0$.  This is complemented by
$\widetilde{F}(0,z)=F(0,z)$.
To extract from $\widetilde{F}(l,z)$ the numbers
$\widetilde{N}(l,p)$, one proceeds as before by Cauchy integration. The
calculation is now much simpler since $(1-2 \xi(z))$ is up to a constant
the jacobian for the change of variables from $z$ to $\xi$. Finally one gets
($l  > 0$):
\be
\label{NtildeRes}
\widetilde{N}(l,p)= \frac{(2d-1)^p\, (2p+l)!}{p!\, (p+l)!}\quad . 
\ee

In some applications, one is interested in a slight variant of the
generating function $\widetilde{F}(l,z)$. Its  definition and final
expression is given by:
\be
\label{Ftildep1Res}
\widetilde{F'}(l,z)\equiv \sum_{p=0}^{\infty} \frac{1}{l+2p}\,
\widetilde{N}(l,p)\, z^p = \frac{1}{l\, (1-\xi(z))^l}
\ee
The last expression, valid for positive $l$, could be obtained  after some work by integration of
$\widetilde{F}(l,z)$. The $ \frac{1}{l+2p}$ in the definition of
$\widetilde{F'}$ occurs naturally when the sum of closed paths is the result
of a fermionic or bosonic determinant. We complement this result with the
one for $l=0$:
\be
\label{tildeF0}
\widetilde{F'}(0,z)\equiv \sum_{p=0}^{\infty}
\frac{1}{2p}\, N(0,p)\, z^p = d\, \log(1 -\xi(z)) + (d-1)\,
\log(1-\frac{2d}{2d-1}\xi(z))\quad .
\ee

The remaining part of this section is dedicated to the evaluation of sums
over simple paths. The basic quantity  is: 
\be
\label{defTTILDE}
\widetilde{{\cal T}}(\mathbf{A}) = \sum_{\bar{L}=0}^{\infty} \ \ \sum_{(n,\vec{\alpha})\in
\tilde{\cal S}_L(n)}
\mathbf{A}_{\alpha_1} \cdots  \mathbf{A}_{\alpha_{\bar{L}}} .
\ee
where $\tilde{\cal S}_L(n)$ is the set of all simple paths with origin in
$n$ and length $L$. In short, what we want is the generalization of the
quantity defined in Eq.~(\ref{defT}) but restricted to simple closed paths. 
Similarly to what we did in Section 3, we will first present the result, and
then give the derivation.
We obtain:
\bea
\label{TTILDEres} 
&\widetilde{{\cal T}}(\mathbf{A})=\frac{1}{(1-\lambda)}\,(2 \lambda(d-1) +(1
+\lambda^2(2d-1))\mathbf{H} -\sum_{\alpha \in I} \mathbf{A}_{\alpha}
\mathbf{H}  \mathbf{A}_{\bar{\alpha}}\\
\label{defH}
&\mbox{    with  }\quad  \mathbf{H}= (1 +(2d-1) \lambda -\tilde{\mathbf{A}})^{-1}
\eea
where $\mathbf{A}_{\alpha} \mathbf{A}_{\bar{\alpha}}=\lambda\, \mathbf{I}$ as
in Section 3. 

The derivation follows a similar track to the one employed for ${\cal
T}(\mathbf{A})$.  Our first goal is the calculation of the  quantity
$ {\cal T}_{\alpha \alpha'}(L, \mathbf{A})$ given by:
\be
{\cal T}_{\alpha \alpha'}(L, \mathbf{A})= \sum_{(n,\vec{\alpha})\in
\tilde{\cal S}_L^{\alpha \alpha'}(n)} \mathbf{A}_{\alpha_1} \cdots
\mathbf{A}_{\alpha_L}\quad ,
\ee
where $\tilde{\cal S}_L^{\alpha \alpha'}(n)$ is the set of all simple paths
$(n,\vec{\alpha})$ of length $L$ ($L > 2$) and origin in $n$ such 
that $\alpha_1=\alpha$ and $\alpha_L=\alpha'$. The main iteration equation 
allowing the evaluation of this quantity is:
\be 
\label{iter1}
{\cal T}_{\alpha \alpha'}(L+2, \mathbf{A})= \sum_{\alpha \ne \bar{\beta};
\alpha' \ne  \bar{\beta'}} \mathbf{A}_{\alpha} {\cal T}_{\beta
\beta'}(L, \mathbf{A}) \mathbf{A}_{\alpha'}
\ee
The sum of ${\cal T}_{\alpha \alpha'}(L, \mathbf{A})$ over $L$ ranging from
$2$ to $\infty$ is denoted ${\cal T}_{\alpha \alpha'}(\mathbf{A})$. An
equation for this quantity follows from summing both sides of
Eq.~(\ref{iter1}) over $L$. After similar manipulations as those of Section
3, one gets: 
\bea
&{\cal T}_{\alpha \alpha'}(\mathbf{A}) = \mathbf{A}_{\alpha} {\cal
T}_{\bar{\alpha}
\bar{\alpha}'}(\mathbf{A}) \mathbf{A}_{\alpha'} + S_{\alpha \alpha'}\\
\nonumber & \mbox{   with  }\  \ S_{\alpha \alpha'}= -\delta_{\alpha \bar{\alpha}'}
\lambda + \lambda \delta_{\alpha \alpha'} \mathbf{A}_{\alpha} +(1+\lambda) 
\mathbf{A}_{\alpha} \mathbf{H} \mathbf{A}_{\alpha'} +\lambda (\mathbf{H}
\mathbf{A}_{\alpha'} + \mathbf{A}_{\alpha} \mathbf{H})
\eea
where $\mathbf{H}$ is the quantity defined in Eq.~(\ref{defH}). Finally
combining the equation for ${\cal T}_{\alpha \alpha'}(\mathbf{A})$
and for  ${\cal T}_{\bar{\alpha} \bar{\alpha}'}(\mathbf{A})$, one can solve
for ${\cal T}_{\alpha \alpha'}(\mathbf{A})$:
\be
\label{resTAAP}
{\cal T}_{\alpha \alpha'}(\mathbf{A}) = \frac{1}{(1-\lambda)}\,
(\lambda(-\delta_{\alpha \bar{\alpha}'} + \lambda \delta_{\alpha \alpha'}
\mathbf{A}_{\alpha}- \mathbf{H}
\mathbf{A}_{\alpha'} - \mathbf{A}_{\alpha} \mathbf{H}) + \mathbf{A}_{\alpha}
\mathbf{H} \mathbf{A}_{\alpha'}+\lambda^2 \mathbf{H}) 
\ee

The previous quantity  can be related to $\widetilde{{\cal T}}(\mathbf{A})$
as follows:
\be
\widetilde{{\cal T}}(\mathbf{A})= \mathbf{I} + \widetilde{\mathbf{A}}
+\sum_{\alpha \ne \bar{\alpha}'}{\cal T}_{\alpha \alpha'}(\mathbf{A})\quad .  
\ee
Using  this result in combination with Eq.~(\ref{resTAAP}) one obtains the final
formula~(\ref{TTILDEres}).  One can again check the validity of  the
expression by using it in reobtaining a known result. We leave this to the
reader. We recall  that in the definition of $\widetilde{{\cal
T}}(\mathbf{A})$ one sums over all simple paths, closed or open. Restricting 
oneself to closed paths can  be done with the same technique explained at
the end of the last section.

\section{Discussion}
In this section we will exemplify how to apply the previous results 
to some physical problems. We consider a lattice model  involving continuous
spin variables with nearest neighbor interactions. These lattice fields can 
be real or complex valued or Grassman variables if they describe fermions. 
To apply the path representation we need a quadratic action or Hamiltonian
in these fields. For example, for complex fields one has:
\be
\label{action1}
\sum_i \phi^a(n)^\dagger \phi^b(m)\, M^{a b}(n,m)\quad . 
\ee
The indices  $n,m$ label  lattice points and the indices $a,b$ are internal.
The matrix $M$ will depend on other fields. For instance, in many cases
constraints or non-quadratic terms in the lattice action can be rewritten as
a quadratic (gaussian) Hamiltonian with the aid of auxiliary fields.
Then, one can integrate out these complex fields ($\phi^a(n)$) using  the Gaussian
integration formulas. The two quantities entering the final expressions are
the determinant of $M$ ($\det M$) and  the inverse of $M$. Now, the nearest
neighbor character of our matrix manifests itself in that we can write 
(after an adequate re-scaling of the fields if necessary):
\be
M= \mathbf{I} - \sum_{\alpha \in I} \Delta_{\alpha} \quad , 
\ee
where the matrix $\Delta_{\alpha}$ can be written as:
\be
\Delta_{\alpha}^{a b}(n,m) = \mA_{\alpha}^{a b}(n)\,  \delta_{m\, n+V(\alpha)}\quad . 
\ee
It only produces transitions between
 a lattice point $n$ and its neighbor in the $\alpha$ direction
$n+V(\alpha)$. It is this form of the matrix $M$ what allows a
random walk (lattice path) representation of the determinant or the inverse
of $M$. Our formulas allow a rearrangement of     this sum over paths 
into a sum over  simple closed paths or terse  paths  respectively.
This is feasible  whenever: 
\be
\Delta_{\alpha} \Delta_{\bar{\alpha}} = \Lambda 
\ee
with $\Lambda$ a matrix which is independent of the lattice point. This
occurs naturally whenever the matrix $\Delta_{\alpha}$, although dependent
on the lattice  point, involves unitary
link fields like in $U(N)$ or $Z_N$ gauge theories. We will restrict  to the case when
$\Lambda$ is a multiple of the identity $\lambda\, \mathbf{I}$. 

Now if we denote by $\mA(\gamma)$ the ordered product of the matrices
$\mA_{\alpha}(n)$ along the path $\gamma$, we can write: 
\bea
\nonumber
&\log(\det(M))/\VOL= \left(-d\, \log(1-\xi(\lambda)) + (d-1)
\, \log(1-\frac{2d}{2d-1}\xi(\lambda))\right) Tr(\mathbf{I}) +\\
&\frac{1}{\VOL}\, \sum_{n \in {\cal L}}\, \sum_{l=1}^{\infty} \frac{1}{l}
\sum_{\tilde{\gamma} \in \tilde{\cal S}_l}(n \rightarrow n)\,  
\frac{Tr(A(\tilde{\gamma}))}{(1-\xi)^l}\quad ,
\eea
where $\VOL$ is the lattice volume and $\xi(\lambda)$ is defined in
Eq.~\ref{xidef}. To arrive to the previous equation,
 we have rearranged as usual the sum over paths into a sum over simple
paths, and used the results of the previous sections. The  term proportional
to $Tr(\mathbf{I})$, equal
to $\widetilde{F}'(0,\lambda)$, gives the contribution of pure spike paths.
In some theories, like $U(N)$ gauge theories at strong coupling 
in the large $N$ limit with
either bosonic or fermionic spin fields in either the fundamental or adjoint
representation, this term turns out to be the only surviving one~\cite{largeN}. Thus, up
to a multiplicative constant depending on the type of field,
$\widetilde{F}'(0,\lambda)$ (Eq.~(\ref{tildeF0})) gives the free energy per unit volume in that
limit. We suggest  that in  other theories, the rearrangement into simple
paths could be an  effective method to perform the summation over paths.    

Now, as an additional application, let us compute the pure spike contribution to the 
{\em mesonic effective potential}. Let us add to the action~(\ref{action1}) a mesonic source 
term:
\be
- \sum_{n,a,b} \phi^a(n)^\dagger \phi^b(n)\, J^{a b}(n)\quad .
\ee
where $J(n)$ acts as the source of local field bilinears (mesons).
Integrating over the gaussian fields $\phi$ we get the connected
generating functional $W(J)$:
\be
W(J)\equiv \log(Z(J)/Z(0))= \sum_{k=1}^{\infty} \frac{1}{k}\, Tr(
(M^{-1}J)^k)\quad ,
\ee
where the trace includes a summation  over lattice points. Each factor of $M^{-1}$
can be expanded into a sum over  paths (random walks). The pure spike
contribution $W_0(J)$ is that in which the overall path obtained within each trace is
a pure spike path. Again this contribution would be the leading one if the
matrices $\mA_{\alpha}(n)$ entering Eq.~(\ref{action1}) involve random $U(N)$
fields at large $N$. In the  subsequent expressions only the remaining part of the
$\mA_{\alpha}$ would enter, which we will assume to be independent of the
lattice point  in what follows.
In order to  implement the  restriction to pure spike paths, it is
convenient to express the propagators $M^{-1}$ as  a sum over terse paths:
\be
(M^{-1}(n,m))^{a b} = \frac{1}{(1-\frac{2d}{2d-1}\xi) }\, 
\sum_{l=0}^{\infty}\, \sum_{\hat{\gamma} \in \bar{\cal S}_l(n \rightarrow m)}\, 
\frac{(A(\hat{\gamma}))^{a b}}{(1-\xi)^l}\quad .
\ee
We then obtain for $W_0(J)$:
\bea
\nonumber
&W_0(J)=\sum_{k=1}^{\infty} \frac{1}{k}\, \sum_{x_1,\ldots ,x_k \in {\cal
L}}\ \ 
 \sum_{\hat{\gamma}_1 \in \bar{\cal S}(x_1 \rightarrow x_2)} \cdots 
  \sum_{\hat{\gamma}_k \in \bar{\cal S}(x_k \rightarrow x_1)} \\
  &Tr(J'(x_1)A'(\hat{\gamma}_1)J'(x_2) \cdots A'(\hat{\gamma}_k))\,  
  \Theta( \hat{\gamma}_1\circ \hat{\gamma}_2 \ldots \circ \hat{\gamma}_k )
\eea
where $\Theta(\gamma)$ is $1$ if $\gamma$ is a pure spike path and zero
otherwise, and the rescaled quantities $\mA', J'$ are given by:
\bea
&\mA'_{\alpha} = \frac{\mA_{\alpha}}{(1-\xi(\lambda))}\\
&J'(n)= \frac{J(n)}{(1-\frac{2d}{2d-1} \xi(\lambda))}\quad .
\eea
We see that the net effect of replacing the sum over paths by a sum over terse
paths is precisely this re-scaling, as  follows from our
results of section 2. The first two terms of $W_0(J)$ are:
\be
\label{linquaterm}
W_0(J)= \sum_{x \in {\cal L}} Tr(J'(x)) + 
\frac{1}{2}  \sum_{x_1, x_2 \in {\cal L}} \overline{J}'(x_1)\, {\cal P}(x_1
\rightarrow x_2)\, J'(x_2) + \ldots 
\ee
The linear term in $J'$ is trivial since the  only path that contributes
is the path of zero length. The constraint $\Theta( \hat{\gamma}_1
\circ \hat{\gamma}_2)$ for the quadratic term  implies that $\hat{\gamma}_2$ must
be the reverse path of $\hat{\gamma}_1$. The resummation over terse paths can
be done with the help of the formulas of section 3. One obtains the following explicit
expression of the propagator ${\cal P}(x_1\rightarrow x_2)$ :
\be
{\cal P}(x_1\rightarrow x_2) = \prod_{\mu} \left(\int_0^{2 \pi}
\frac{d\varphi_{\mu}}{2 \pi}\,
e^{\imath \varphi_{\mu}(x_1-x_2)_{\mu}}\right)\, (1-\lambda')\, (1 +(2d-1)\lambda'
-\mathbf{B})^{-1}  
\ee
where:
\bea
&\lambda'=\frac{\lambda^2}{(1-\xi(\lambda))^4}\\
&\mathbf{B}=\sum_{\alpha \in \In} e^{\imath \varphi_{\alpha}} \mA'_{\alpha} 
\otimes (\mA'_{\bar{\alpha}})^t
\eea
These expressions were used in our recent paper on $N=1$ SUSY
Yang-Mills~\cite{GGAP}. In formula~(\ref{linquaterm}) $J'(n)$ has to be
looked at as a column vector on which  the matrix ${\cal P}$ acts. Then,
$\overline{J}'(x_1)$ is the row vector whose elements are the transpose
of $J'$.

Finally, we will address the calculation of the cubic term in $W_0(J)$.
For that purpose we have to solve the constraint  imposd by $\Theta( \hat{\gamma}_1\circ
\hat{\gamma}_2 \circ  \hat{\gamma}_3)$. In the generic case, it can be solved as follows:
\bea
&\hat{\gamma}_1 = s_2^{\alpha}\circ  (s_3^{\beta})^{-1}\\
&\hat{\gamma}_2 = s_3^{\beta}\circ  (s_1^{\gamma})^{-1}\\
&\hat{\gamma}_3 = s_1^{\gamma}\circ  (s_2^{\alpha})^{-1}
\eea
where $s_1^{\gamma} \in \bar{\cal S}^{\gamma}$ is a terse path ending with a step in the $\gamma$
direction, and similar definitions apply for $s_2$ and $s_3$. Furthermore, one
must have $\alpha \ne \beta \ne \gamma \ne \alpha$. The exceptional cases
occur when any of the paths $s_i$ is
a path of zero length.  It is clear that the summation over the paths $s_i$
can be done with the aid of the formulas of section 3. The best way to
express the result is in terms of the mean mesonic field $\Phi^{a b}(x)$:
\be
\Phi(x) = \sum_{x' \in {\cal L}} {\cal P}(x \rightarrow x')\, J'(x')  
\ee
Then the cubic term in $W_0(J)$ becomes:
\be
\label{Wcubic}
\sum_{x \in {\cal L}} \left( \frac{1}{3}\, Tr(\Phi^3(x)) - \sum_{\alpha \in
\In} Tr(\Phi(x) \Phi_{\alpha}^2(x)) +  \frac{2}{3}\, \sum_{\alpha \in\In}
Tr(\Phi_{\alpha}^3(x)) \right) \quad ,
\ee
where:
\bea
&\Phi_{\alpha}(x)=\frac{1}{1-\lambda'}\, (\mA'_{\alpha} \Phi(x
+V(\alpha)) \mA'_{\bar{\alpha}} -\lambda' \Phi(x))\, =\\
&\sum_{y \in {\cal L }}\, \sum_{l=1}^{\infty}\, \sum_{\hat{\gamma} \in \bar{\cal S}^{\alpha}_l(x \rightarrow
y)}\,
A'(\hat{\gamma})\,J'(y)\, A'(\hat{\gamma}^{-1}) \quad .
\eea
The effective action $\Gamma_0(\Phi)$ can be obtained by a Legendre
transformation from $W_0(J)$. The cubic term is precisely given by minus the
corresponding cubic term in $W_0(J)$, displayed in Eq.~(\ref{Wcubic}). 
Following a similar procedure one can  use the formulas of the previous
sections to compute quartic and higher
terms in  $\Gamma_0(\Phi)$.

\section*{Acknowledgements}
The author acknowledges useful conversations with E. Gabrielli and C. Pena. 
This work has been partly  financed by CICYT under grant AEN97-1678.

\end {document}